# Improving Collaborations in Neuroscientist Community

Isabelle Mirbel and Pierre Crescenzo
*Université de Nice Sophia-Antipolis / Laboratoire I3S*
*930 route des Colles – BP 145*
*F-06903 Sophia-Antipolis cedex, France*

**ABSTRACT**

In this paper, we present our approach, called SATIS (Semantically AnnotaTed Intentions for Services), relying on intentional process modeling and semantic web technologies and models, to assist collaboration among the members of a neurosciences community. The main expected result of this work is to derive and share semantic web service specifications from a neuroscientists point of view in order to operationalise image analysis pipelines with web services.

## 1. INTRODUCTION

Computational neurosciences aim at analyzing neurological datasets and studying brain functions. The analysis of users' image processing pipelines shows many commonalities in data sets and processing chains. The manipulated data are mostly images completed with clinical information and additional annotations. As highlighted in [1], basic processing, as for instance intensity corrections or tissue classifications, are common to several image analysis pipelines, while each pipeline also contains specific processing such as brain structure segmentations or image interpretations. In practise, there is no sharing of common basic processing units among the different processing chains. In this context, web services appear to be a privileged mean to support dedicated processing pipelines for each targeted application and to share common basic processing units inside a neuroscientist community.

To facilitate the exploitation of web resources (documents, actors or services), the semantic web research community aims at making explicit the knowledge contained into resources. This knowledge is represented by ontologies which structure terms, concepts and relationships of a given domain. Ontologies are often used to extract and represent the meaning of resources. This meaning is expressed through annotations supporting semantic resources indexing in order to explicit and formalise their content. Resource retrieval inside the community relies on the formal manipulation of these annotations and is guided by ontologies.

As it is shown in figure 1, our work takes place in the context of a community of neuroscientists building image processing pipelines for their targeted application and therefore relies on web services (from their own registry or from a web registry). Web services are annotated by meta-data supporting their manipulation. But when the number of web services becomes important in the community registry, it may be difficult for neuroscientists to be aware of available web services. It may also be difficult for each neuroscientist to rely on web services provided by other neuroscientists while building image processing pipelines. Moreover, a registry of web services annotated by meta-data is not enough to support image processing pipelines operationalization by non computer scientists. Additional support is required to help them to understand how available web services meet their needs.

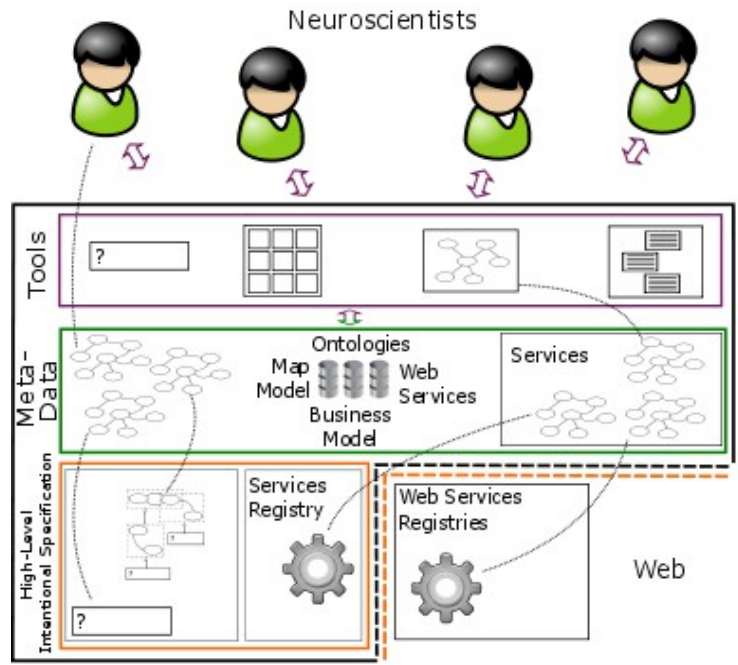

**Figure 1 : Neuroscientists community**

Therefore, we propose SATIS an approach to specify high-level business-oriented activities with the help of an intentional model and to derive web services specification from this high-level description. As one of our aim is to improve collaboration and sharing inside the community, we also propose to consider high-level intentional specification of processing pipelines as resources of the community. Therefore, we provide means to annotate high-level intentional specification in order to assist their retrieval and sharing among the neuroscientists of the community.

## 2. RELATED WORK

Our work takes place in the web services domain and concentrates more specifically on service discovery and selection. It belongs to the family of goal-based service retrieval approaches. These approaches ([2],[3],[4],[5]) aim at specifying the goals which have to be satisfied by the retrieved services. In these proposals, different models are proposed to specify goals but none addresses the problem of how to capture goals. They all consider that goals have already been identified and specified. On the contrary, our aim is to provide means to assist neuroscientists in querying the web services registry to find web services to operationalise a processing pipeline; and we are particularly interested in providing means to elicit and specify neuroscientists requirements in terms of services, upstream of the previously cited approaches.

The approach discussed in [6] also addresses this issue by proposing models and tools to capture user's goals with the help of an ontology or in natural language. What sets us apart from this approach is that we propose an incremental process to refine neuroscientists' requirements in order to specify the features required for the web services under retrieval. In [6], sentences are mapped into a set of concepts and relationships on which is based the web service retrieval process. There is no goal decomposition support. On the contrary, our approach focuses on intention refinement in order to support goal decomposition and to improve sharing and reuse of user requirements at different levels of specification.

The approach presented in [7] proposes a move towards a description of services in business terms. A requirements engineering process to determine intentional services that meet business goals is also proposed. Our approach distinguishes itself from [7] by the fact that we rely on semantic annotations and semantic web models and techniques to enrich the goal (or intention) specification, in order to provide reasoning and explanation capabilities. Moreover, we focus on specification sharing and reuse as well as web service discovery while in [7] the focus is on guiding the elicitation of technical specifications from business ones.

In [8], a collaborative tagging system to improve web service discovery is proposed in order to offer an alternative in domains where there are many resources or contents to classify. In this approach, tags are considered as keywords and combiners are provided in order to write query to retrieve web services. When this approach aims at improving web service descriptions by relying on users vocabulary, we do concentrate on user intentions modeling and bindings between web services and user tasks. Approaches relying on past users'experiences [9], user preferences [10],[11] have also been provided. On the contrary of these approaches, in our work, the focus is not on the user characteristics but on the user tasks that we aim at modeling and from which web service characterics are derived.

With regards to approaches dealing with ontology-based service discovery [12], and more precisely OWL-S[13] based approaches (as we are relying on OWL-S with regards to Web Service descriptions), capability matching and matchmaking algorithms, mainly exploiting subsumption relationships, as well as ranking mechanisms have also been proposed [14][15][16]. Our approach distinguishes itself from these works by the fact that our focus is on providing means to assist final users in authoring queries and not rendering them. We are indeed interested in the upstream process of deriving queries from final users requirements and in providing support to annotate such queries in order to enhance their capitalisation and sharing among the community members.

Beyond an alternative way to discover and retrieve web services, we also provide means to capitalise know-how about web service discovery and search processes themselves. Another novelty of our approach is to operationalise goals by rules in order to promote both sharing of high-level intentional specification and cross fertilisation of know-how about web services discovery and search processes among the community members.

The paper is organised as follows. In section 3, we discuss the different collaboration means provided in our approach. In section 4, we detail SATIS elicitation step relying on an intentional process model. In section 5, we explain how semantic annotations and queries are derived from image analysis pipelines in order to enhance knowledge sharing among the community members. Then in section 6, we discuss intentional fragmentation of know-how about how to operationalise image analysis pipeline before explaining how the SATIS approach supports web services discovery in section 7. Then, in section 8, we conclude and give some perspectives of our work.

## 3. SATIS APPROACH

By providing support to web service discovery and retrieval for non-computer scientist users, our main objective is to promote know-how sharing among community members. More specifically, we assist the know-how *transfer* from expert members to novice ones by providing means to incrementally specify high-level business-oriented activities with an intentional modeling technique. Indeed, we provide means to populate a library of high-level intentions defined at different abstraction levels and allowing a novice user to start his/her web services discovery and search process at the level of specification s/he is comfortable with. S/he is then guided by the know-how previously entered by expert members into the community memory to derive from the initial requirements a set of web services (or basic process units) specification.

As it has been previously explained, beyond a way to discover and retrieve web services, our approach aims at providing means to promote sharing of high-level intentional specification and cross fertilisation of know-how about search processes among the community members. Indeed, our second objective is to assist the know-how *sharing* among expert members. Therefore, high-level incremental specifications of intentions are decomposed into fragments, highlighting the reusable dimension of high-level specifications out of the scope of the targeted application. We also propose an operationalisation of fragments by rules to take advantage of inference capabilities to discover alternative know-how to operationalise a web service discovery and search process. When a neuroscientist searches for web services to opertionalise an image processing pipeline, s/he provides high-level intentional requirements or select some of them in the community repository. S/he may refine the high-level requirements into more precise intentional requirements until the decomposition level allows to associate a query to search for web services to part of the image processing pipeline. During the refinement task, the neuroscientist may provide his/her way of decomposing the image processing pipeline or select fragments already stored in the community memory. The rule base implementation of fragment promotes the reuse of existing fragments. Indeed, reuse of pipeline parts is enhanced by their decomposition into fragments. Moreover, fragments elicited from different image processing pipelines may be reused when operationalising a new image processing pipeline, thus contributing to interchange between different way of thinking image processing pipeline operationalisation and so supporting cross fertilisation inside the scope of the community.

Finally, in such communities, web services are provided by computer scientists and high-level intentional specifications are specified by neuroscientists. In addition to assist know-how transfer between novices and experts and to share know-how between experts, our aim is to support *collaboration* between service providers (computer scientists) and service consumers (neuroscientists). Indeed, our last objective is to provide both :
- means for service consumers to identify and specify their requirements and transmit them to service providers and

- means for service providers to disseminate information about available services.

By relying on a rule based specification to derive web services specification and by providing distinct and dedicated modeling techniques to both service providers and service consumers as well as mapping mechanisms between them, we assist the bidirectional collaboration between neuroscientists and computer scientists.

To support the different collaboration means discussed above, we propose an approach based on: the map formalism [17] to identify and specify high-level intentional specification. And we rely on the W3C standards RDF, RDFS and SPARQL to provide means to define a common vocabulary, to annotate both web services (i.e. basic process units) and intentions, to query the intention library as well as the service registry, and to reason on them.

The map model was introduced in the information system engineering domain to model processes on a flexible way. According to [18], a map is a process model in which a non-deterministic ordering of intentions and strategies has been captured. A map is a labeled directed graph with intentions as nodes and strategies as edges between intentions. An intention is a goal that can be achieved by following a strategy. An intention expresses what is wanted, a state or a result that is expected to be reached disregarding considerations about who, when and where. There are two distinct intentions that represent the intentions to start and to stop the process. A strategy characterizes the flow from the source intention to the target intention and the way the target intention can be achieved. Indeed, a map contains a finite number of paths from its start intention to its stop intention, each of them prescribing a way to achieve the goal of the process under consideration. Compared to other process modeling formalisms [19] the map model captures not only how a process proceeds but also why by (i) focusing on process goals (intentions) instead of process activities or process results and (ii) embedding contextual information. It also support different levels of abstraction thus facilitating sharing and reuse of modeled processes.

By relying on W3C standards RDF for data interchange on the Web and RDFS to name and define a vocabulary to be used in RDF annotation graph, we take advantage of existing domain ontologies as well as proposal to semantically annotate web services. SPARQL provides a query language for RDF graphs, a language results to represent the answers to a query and a protocol to submit a request to a remote server and receive responses. By relying on this W3C standard, we take advantage of semantic search engines, like CORESE [21] for instance, which enables the processing of RDFS and RDF statements and also perform SPARQL queries and run rules over the RDF graph. In this context and with regards to web service annotation, our approach is not dedicated to a particular kind of semantic meta-data. In the future, we wish to propose a framework enhancing web service discovery regardless of the ontology or technique used to annotate web services. In this paper and as a first step, we illustrate our approach with web services annotated using the OWL-S ontology [13]. In this upper ontology for services, the profile provides the information needed to discover a service, the model and the grounding, taken together, provide information to make use of a service. In our work, we rely on the profile and grounding parts of the OWL-S description as well as the description of inputs and outputs in the process part.

Our approach aims at providing to neuroscientists which are not familiar with computer science, a complete solution to easily use a set of web services. Our approach is decomposed in four steps:
- image analysis pipelines *elicitation* aiming at capturing users requirement and operationalisation means,
- image analysis pipelines *annotation* aiming at associating meta-data to users requirement and operationalisation means in order to enhance their sharing among the community members,
- image analysis pipelines *fragmentation,* consisting in breaking users requirement and operationalisation means into self-governing pieces in order to support their reuse,
- image analysis pipelines operationalisation consisting in relying on already stored users requirement and operationalisation means to search for web services to operationalise an image processong pipeline.

The approach is further detailed in the following sections.

## 4. INTENTIONAL ELICITATION

The focus of the elicitation step of our approach is to capture know-how about image analysis pipelines in order to support reuse and sharing about how to operationalise such pipelines. In other words, we are interested in know-how about how to search for web services in order to support image analysis pipelines operationalisation.

We define search processes we are interested in as sequences of atomic searches to be processed by a neuroscientist to fulfill an image analysis pipeline. A search process may be seen as a particular kind of business process limited to search activities. Different business process modeling formalisms have been proposed in the literature. They can be classified into three categories: activity-oriented, product-oriented and decision-oriented ones [19]. Decision-oriented models are semantically more powerful than the two others because they explain not only how the process proceeds but also why. Their enactment guide the decision making process that shapes the process, and helps reasoning about the rationale [19]. To support knowledge transfer about search process from experts to novices, we are concerned with the modeling of why the search process is decomposed the way it is, as well as with the specification of how it is decomposed. Moreover, to handle different users'profiles and levels of knowledge, we want to provide means to specify search processes at different levels of detail. For all these reasons, we propose to model search processes by adapting an intentional process modeling formalism : the map model [17][18]. We gathered the concepts and relationships of the map model and we built an RDFS ontology dedicated to the representation of intentional processes [20].

During the first step of SATIS approach, dedicated to elicitation, final users (neuroscientists) define their image analysis pipeline by describing intermediate *intentions* (i.e. goals and subgoals to be satisfied through the processing chain) and *strategies* (i.e. means to reach goals). Figure 2 gives an example of two intentional maps. The map on the upper part of the figure contains three intentions defined by a neuroscientist : *image pre-processing, skull striping* and *image segmentation*. Between the intentions, we discover strategies. Strategies define the way to pass from an intention to a next one. There can be many strategies which link up the same intentions.

In this case, a label is associated to each strategy in order to elicit its particular features. An image analysis algorithm is an example of meaningful strategy when different algorithms exist to transform an image.

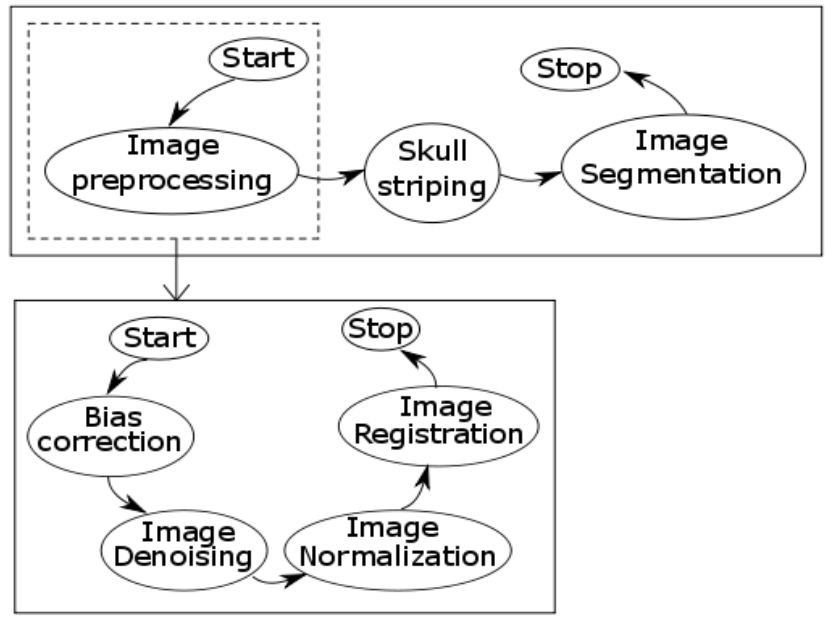

**Figure 2 : Intentional representation of image analysis pipeline**

To further formalize intentions and strategies, we rely on [22] proposal, which has already proven to be useful to formalize goals [23][24]. According to [22], an intention statement is characterized by a verb and some parameters which play specific roles with respect to the verb. Among these parameters, the main one is the object on which the action described by the verb is processed. Let us consider again the map depicted in figure 2. Intention *image preprocessing,* for instance, is described by its verb *preprocessing* and its object *image.*

Indeed, in a map, each set which is made up by a source intention, a strategy and a target intention is a section of the map. In figure 2, an example of section has been highlighted by a dotted line. A section can be refined by giving a new map which describes how to reach the target intention in a more detailed way (by using more specific and low-level intentions and strategies). The map presented in the bottom of figure 2 is a refinement of the section highlighted with a dotted line in the map of the upper part of figure 2. Indeed, the second map details how to do image preprocessing : by doing *bias correction*, then *image denoising, image normalization* and *image registration*. Map sections are refined by more detailed maps until it is possible to associate to each map section a query to search for web services to operationalise the section under consideration.

Let's precise that an intentional map is neither a state diagram, because there is no data structure, no object, and no assigned value, nor an activity diagram, because there is always a strong context for each section of the map: its source intention and its strategy. We can attach more information to this kind of schema (in order to help the user of the map to choose the adequate strategy, for example), but this is not the goal of this paper to fully describe the map model.

So at the end of this step, we obtain a set of maps organized through refinement relationships where each section not refined by a map is operationalised by a query eliciting how to search for web services to achieve its target intention possibly from a source intention and following a particular strategy if it is specified.

The map model has proven to be suitable to capture user requirements in several application domains [23][25][26]. By relying on it, we aim at supporting the elicitation of know-how about how to search for web services to operationalise image analysis pipelines as well as the sharing and reuse of this know-how inside a community of neuroscientists..

## 5. SEMANTIC ANNOTATION

In the context of a neuroscientists community, the objective of our work is to enhance knowledge sharing through the development of dedicated knowledge management services. Knowledge management services aim at offering efficient and effective management of the community knowledge resources. In our case, the knowledge resources we are interested in are maps about image analysis pipelines, queries about how to search for web services and web service descriptions. To achieve efficient knowledge sharing through the development of knowledge management services, we rely on ontologies and on semantic annotations of the community knowledge resources with regard to these ontologies.

In SATIS, we adopt web semantic languages and models as a unified framework to deal with image analysis pipelines specification and web service descriptions themselves. We gathered the map model concepts and relationships into an RDFS ontology dedicated to the representation of intentional processes: the map ontology [20]. As a result, intentional processes annotated with concepts and relationships from this ontology can be shared and exploited by reasoning on their representations. We also consider semantic Web Service descriptions specified with the help of the OWL-S ontology. And queries to search for available web services to operationalise image analysis pipelines are specified with the help of the W3C standard query language for RDF annotations: SPARQL. Our approach relies on three ontologies: The map ontology we proposed, the OWL-S ontology and a domain ontology (in our case an ontology describing medical images and medical image processing dedicated to the neuroscience domain). The ontology maintenance as well as the mamangement of multiple users domain konwledge is out of the scope of this work in which we rely on an existing domain ontology. With regards to knowledge about intentional users requirement and their operationalisation means, they are stored in the community memory and provided to the neuroscientists when requested. If several operationalisation means exist to search for web services to operationalise an image processing pipeline, they are provided to the neuroscientist as alternative know-how to answer his/her need.

The semantic annotation step of SATIS approach consists in generating and/or writing (this is designed to be a semi-automatic transformation) both semantic annotations about map sections and queries to search for adequate web service or set of web service specifications in order to operationalise image analysis pipelines. Indeed, each section of a map may be refined by (i) a map providing more details about how to achieve the target intention from the source intention

(possibly by following a particular strategy) and/or (ii) a query to be run to search for web services supporting the operationalisation of the map section under consideration.

Examples of translation of map sections into semantic annotations are given in figure 3. In these examples, where namespace *map* refers to the map ontology and namespace *dom* refers to the domain ontology, sections are specified by a sources intention and a target intention (no particular strategy is defined). Intentions are described through the domain concepts *Preprocessing, SkullStriping, Segmentation* which are instantiating concept *Verb* of the map ontology and *Image* which is instantiating concept *Object* of the map ontology.

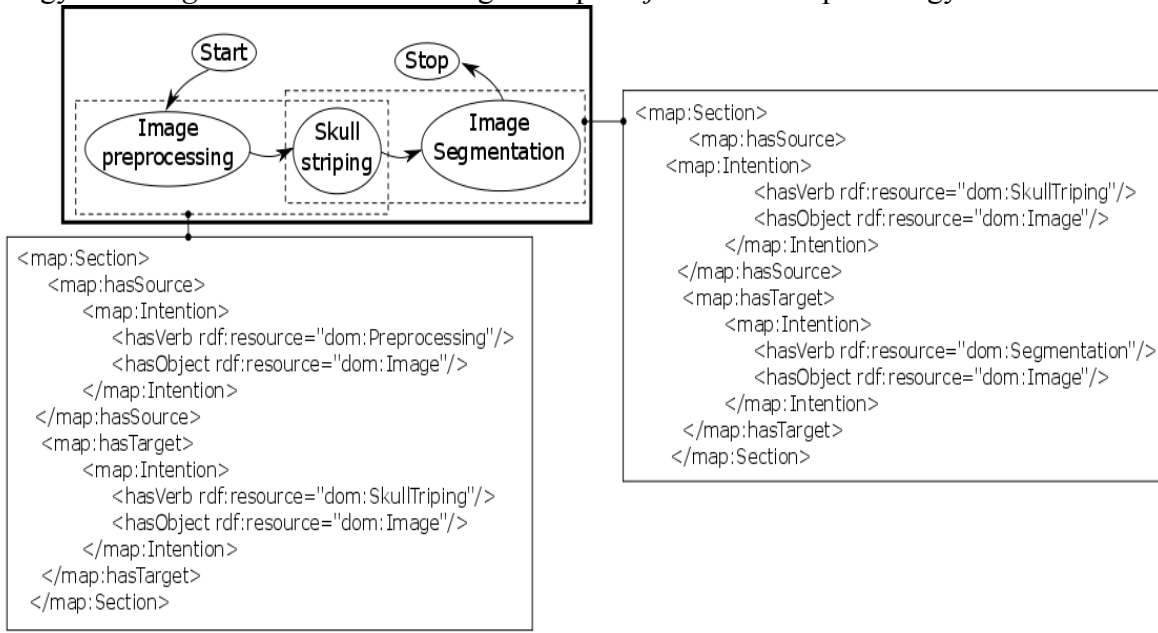

**Figure 3 : Example of map semantic annotation**

Examples of queries to search for web services to operationalise map sections are given in figure 4. In these examples, where namespace *process* refers to the OWL-S process ontology, the input and output parameters of the available web services are exploited to select the right web service descriptions in order to operationalise the *image preprocessing* and the *image segmentation* parts of the image analysis pipeline under consideration.

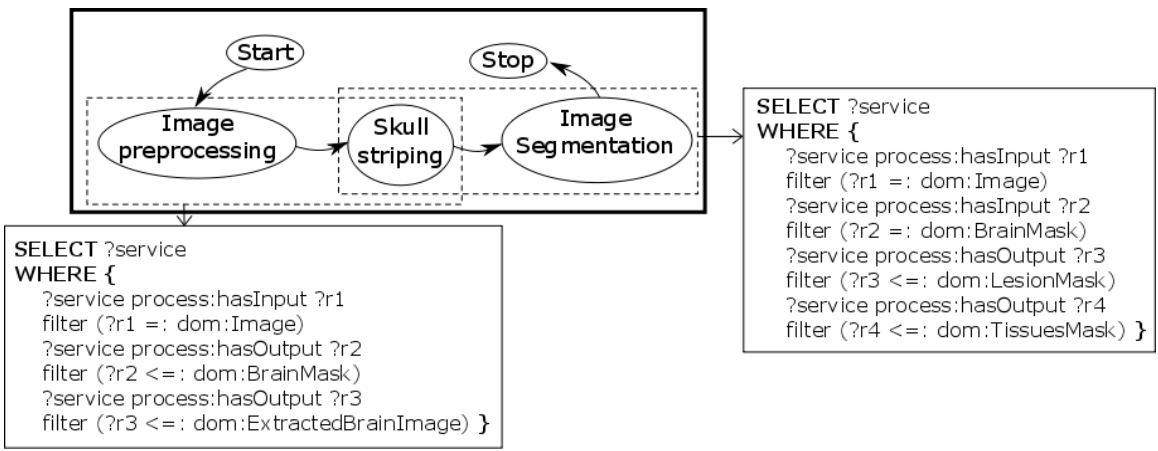

**Figure 4 : Example of queries to operationalise maps**

By associating queries to map sections (instead of web service descriptions themselves), we assume a loosely coupling between image analysis pipelines on one hand and web services descriptions on the other hand: if new web service descriptions are added inside the web service registry, they can be retrieved to operationalise an image analysis pipeline even if it has been specified before the availability of the web services under consideration; and if web services are removed from the web service registry, the image analysis pipelines that they satisfied are still valid and may be operationalised by other available web services. Indeed, web services are dynamically selected when rendering queries associated to map sections.

Semantic annotations about image analysis pipelines and web services can then be used by knowledge management services such as knowledge search services, knowledge visualization services and therefore support the sharing processes in the neuroscientists community.

## 6. INTENTIONAL FRAGMENTATION

In the third step of our approach, we aggregate all specifications captured during the semantic annotation phase into fragments in order to promote among the community members both sharing of image analysis pipeline specifications and cross fertilisation of know-how about how to search for web services.

Indeed, a fragment aims at providing reusable means to operationalise part of an image analysis pipeline. Therefore, it embodies a map section as signature and an operationalisation means as body. Map sections may be operationalised by a more detailed map or a query. So, we distinguish two kinds of fragments: Intentional fragments providing maps and operational fragments providing queries.

A fragment is represented by a rule which conclusion represents a section of a map and which premise represents either an operational means (a query) or an intentional means (a map) allowing to achieve the target intention of the section in conclusion. We call a rule concrete or abstract depending on wether its premise represents operational or intentional means.

The SPARQL language provides a unified framework to represent both concrete and abstract rules through the CONSTRUCT query form. A CONSTRUCT query form returns an RDF graph specified by a graph template and constructed by taking each query solution, substituting for the variables in the graph template and combining the resulting RDF triples. In our case we formalize a rule representing a fragment by a SPARQL query. Its CONSTRUCT clause is the conclusion of the rule, i.e. the graph template to construct the RDF representation of a section of a map. Its WHERE clause is the premise of the rule, i.e. a graph pattern representing a map (abstract rule) or criteria for retrieving relevant web service descriptions (concrete rule). During this aggregation of maps and queries into rules, the original intentions and strategies are naturally modularised and this fact far improves the reusability of the concerned search process.

### 6.1 OPERATIONAL FRAGMENTS

Figure 5 shows an example of operational fragment. On the left side of the figure, a graphical illustration of the fragment is presented while its corresponding rule implemented by a SPARQL CONSTRUCT query form is shown on the right side of the figure. This fragment has been extracted from the image analysis pipeline of figure 2. Since the query provided in the fragment body (left bottom side of figure 5) does not assume any pre condition on the processed image (*dom:Image* as input) and no particular algorithm is specified, only the target intention has been specified in the fragment signature. No source intention (i.e. pre condition) and no strategy (i.e. manner to achieve the target intention) have been specified in the fragment signature.

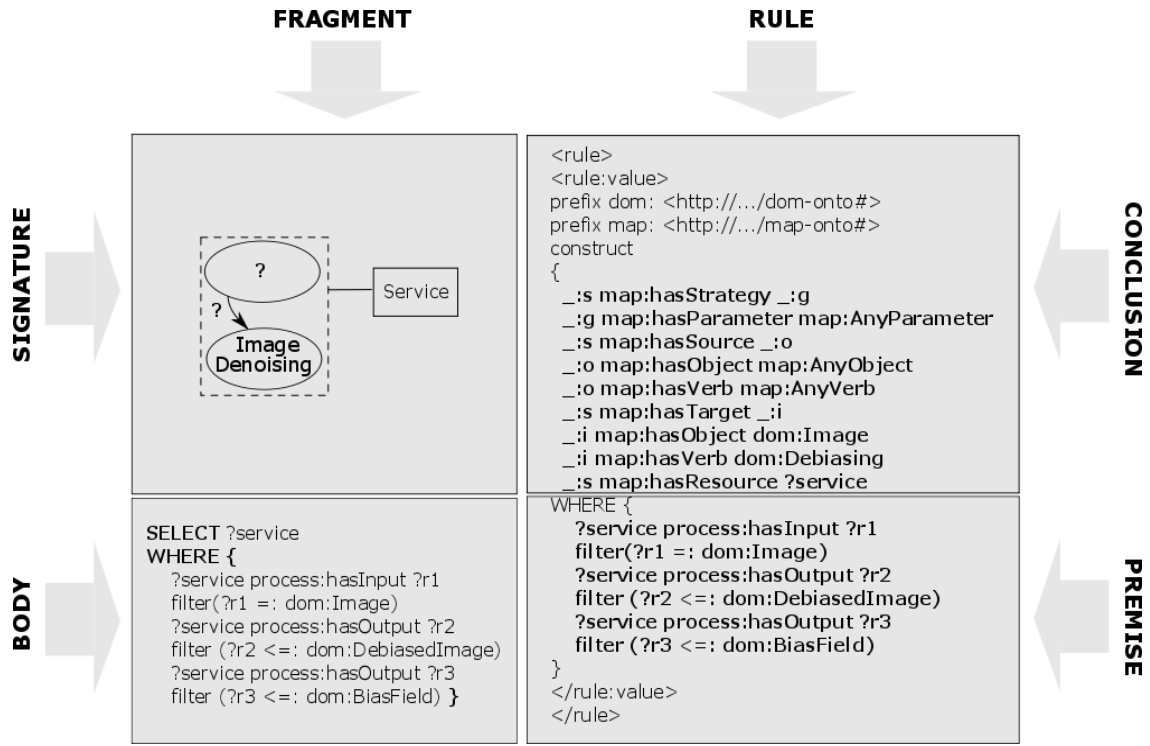

**Figure 5 : Example of operational fragment**

On the right side of figure 5, the CONSTRUCT clause of the rule implementing the fragment under consideration is a graph template for building an RDF graph representing any map section aiming at searching for web services descriptions about image debiasing. It includes both statements describing the target intention of the section with the domain concepts *Image* and *Debiasing* respectively instantiating concept *Object* and *Verb* of the map ontology and statements about the RDF graph pattern to search for web services operationalising the section and which content is described in the WHERE clause of the query. This links together the intentional and operational levels. The WHERE clause of the query describes how to operationalise any section (in particular the one of our example) whose RDF representation matches with the graph template in the CONSTRUCT clause. It is a graph pattern that matches with the RDF web service descriptions. It includes statements about the input and output of the web services.

## 6.2 INTENTIONAL FRAGMENTS

A map section operationalised by a more detailed map is indeed represented by a couple of fragments implemented by a couple of rules.

One fragment embodies the map section to be operationalised as signature and the detailed map as body. An example of such a fragment is shown in figure 6.

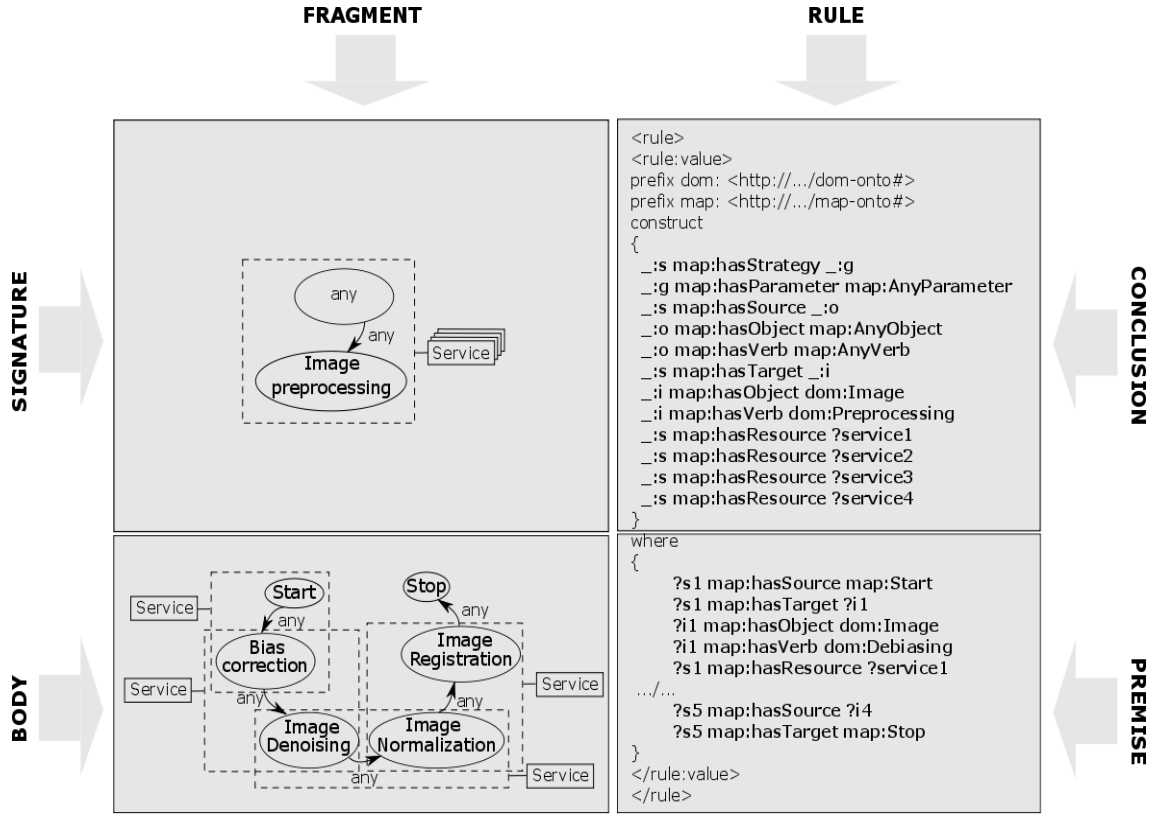

**Figure 6: Example of intentional fragment (part 1)**

This fragment implements the operationalisation of the section highlighted in figure 2. Its body corresponds to the map shown on the bottom part of figure 2. The CONSTRUCT clause of the rule is a graph template for building an RDF graph representing any map section aiming at searching for web services descriptions about image preprocessing. It includes both statements describing the target intention of the section with the domain concepts *Image* and *Preprocessing* respectively instantiating concept *Object* and *Verb* of the map ontology and statements about the RDF graph operationalising the section and which content is described in the WHERE clause of the query. This links together the two levels of intention refinement. The WHERE clause of the query describes how to operationalise any section (in particular the one of our example) whose RDF representation matches with the graph template in the CONSTRUCT clause. It is a graph template that matches with the RDF representation of the map shown in the bottom part of figure 2. It includes statements about five sections: the first ones describe a first section *?s1* which source intention is a *start,* which target intention has for object *image* and for verb *debiasing* and which is operationalisable by the web services *?service1*; the following ones describe a second section  *?s2* which source intention is the target intention of the first section *?s1,* which target intention has for object *image* and for verb *denoising* and which is operationalisable by the web services *?service2* and so on. The retrieved services are also part of the graph template in the CONSTRUCT clause in order to be propagated all along the search process. For readability purpose, only part of the graph template of the WHERE clause is shown in figures 6 and 7.

In SATIS, searching for web services by relying on operational and intentional fragments is achieved by applying rules implementing fragments in backward chaining. As operational fragments publish section as signature (rule conclusion), when embedding intentional operationalisation means into fragments, in addition to a first fragment associating a section and its more detailed map, we need a second fragment specifying how to build the more detailed map from a set of sections obtained by applying operational rules in backward chaining. Figure 7 shows the fragment allowing to build the map shown in the fragment body of figure 6.

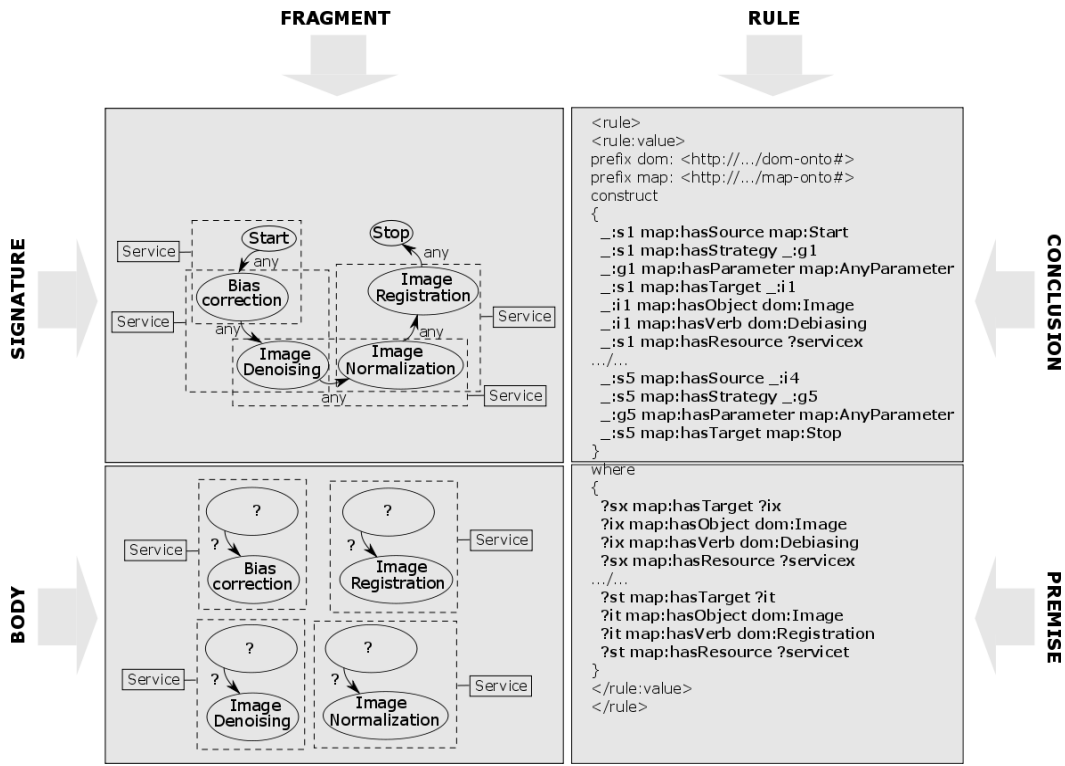

Figure 7 : Example of intentional fragment (part 2)

The CONSTRUCT clause of the rule implementing the fragment under consideration is still a graph template. But in this case it aims at building an RDF graph representing any map aiming at searching for web services descriptions about image debiasing, image denoising, image normalization and image registration. It includes statements about the five sections required to do image preprocessing and statements about the RDF graphs operationalising the different required sections and which contents are described in the WHERE clause of the query. The WHERE clause is a set of five graph templates that matches with the RDF representation of the five sections of the map shown in the bottom part of figure 2. It includes statements about five sections: the first ones describe a section *?ix* which target intention has for object *image* and for verb *debiasing* and which is operationalisable by the web services *?servicex*; the last ones describe a section *?st* which target intention has for object *image* and for verb *registration* and which is operationalisable by the web services *?servicet.*

Thanks to this couple of fragments, web services retrieved with the help of operational fragments may be agregated into maps in order to operationalise more high level sections and so on until the whole image analysis process is operationalised.

At the end of the intentional fragmentation step, the community memory has been enriched by a set of intentional and operational fragments extracted from the image analysis pipelines elicited in the first step of the SATIS approach. This fragment repository aims at improving know-how sharing and cross fertilisation of means to operationalise image analysis pipelines as it will be explained in the followind section.

## 7. BRIDGING THE GAP BETWEEN USER'S NEED AND WEB SERVICES

The last step of SATIS consists in searching web services specifications to operationalise a set of intentions and strategies associated to an image analysis pipeline. We rely on a semantic search engine like CORESE [21] for both i) backward chaining on the knowledge base of rules implemented as SPARQL CONSTRUCT query form and ii) matching with the knowledge base of RDF annotations describing available web services. The knowledge base only stores the queries, and not the maps. These are dynamically created when needed all along the backward chaining process, as temporarily subgoals, until web services annotations are found to match all the subgoals and therefore the general goal section. As a result, a neuroscientist searching for means to operationalise an image analysis pipeline will take advantage of all the rules and all the web service annotations stored in the community memory at the time of his/her search. This memory may evolve over the time and therefore the web services retrieved by applying a rule may vary as well. In other words, the association of web service descriptions to map sections (i.e. parts of an image analysys pipeline) is done at runtime and depends on the web services available in the community memory.

The upper part of figure 8 shows an example of map dealing with another image analysis pipeline in which the first section has been refined into a more precise map shown in the bottom of the figure. Following the SATIS approach, these maps are completed by queries in the first step of the approach. Then semantic annotations and SPARQL queries are generated and intentional and operational fragments are derived. Among the fragments added to the community memory, a couple of them are dedicated to the section highlighted in the map of the upper part of figure 8. One of this two fragment has the same signature as the fragment presented in figure 6, when the other fragment provides an alternative way to implement the section highlighted in figure 8, as the fragments presented in figures 6 and 7 already provide a way to operationalise the section under consideration.

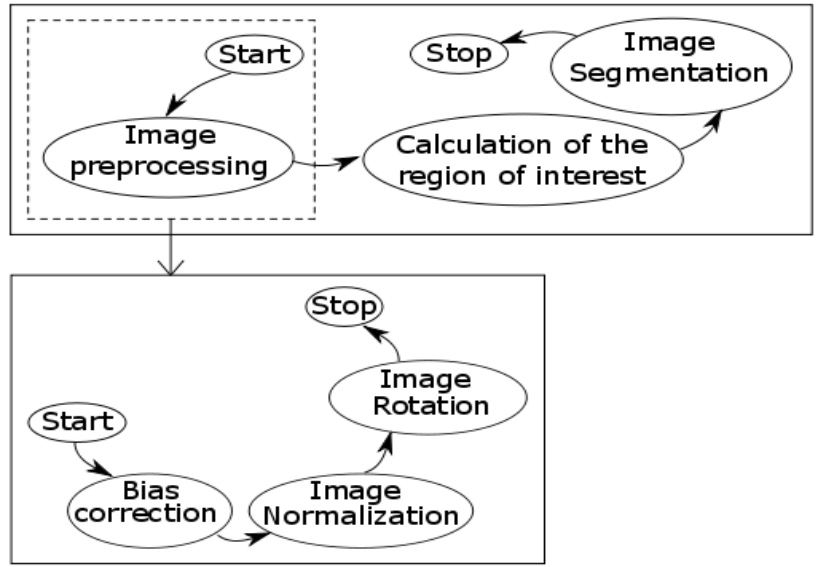
**Figure 8 : Example of map**

When a neuroscientist interested by the map shown in the upper part of figure 2 will search for web services to implement this image analysis pipeline inside the framework of the SATIS approach, s/he will take advantage of all the rules and all the web service annotations stored in the community memory. Two ways to do image pre-processing (the one shown in figure 2 and the one shown in figure 8) will be exploited. Concrete rules implementing operational fragments dealing with bias correction, image denoising, image normalization, image registration and image rotation will be exploited through the backward chaining mechanism. If web services annotations are retrieved (i.e. web services annotations match the graph templates of WHERE clauses of concrete rules) the corresponding sections will be dynamically constructed as subgoals of the image analysis pipeline under operationalisation. If at least one web service annotation match each graph template of each concrete rule corresponding to our running example, then two ways to operationalise the image pre-processing stage of the image analysis pipeline will be provided to the neuroscientist, illustrating a case of cross fertilisation of know-how about how to search for web services, as shown in figure 9.

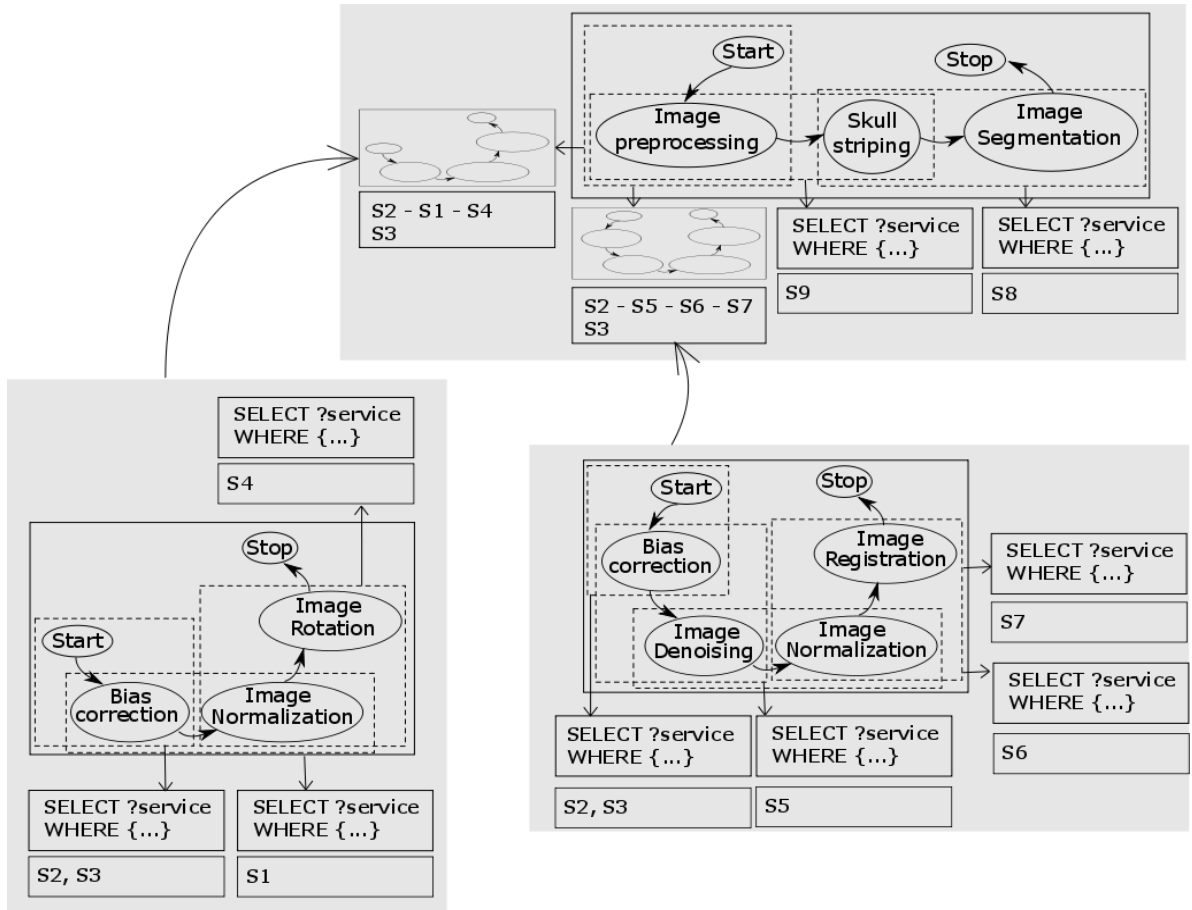

**Figure 9 : Example of cross fertilisation**

Let's clarify that the result is composed by specifications of candidate web services, and not by web services themselves. In the example of figure 9, different sets of web services will implement the whole image analysis pipeline. The invocation of the selected (among the candidates) web services is done dynamically and is out of the scope of this work.

## 8. CONCLUSION

In this paper, we presented SATIS, an approach, relying on intentional process modeling and semantic web technologies and models, to assist collaboration among the members of a neurosciences community. Our main objectives are more precisely to assist the know-how transfer from expert members to novice ones, to promote cross fertilisation of know-how among community members and to support collaboration between computer scientists and neuroscientists. Therefore, starting from an intention based image analysis pipelines elicitation, we adopt web semantic languages and models as a unified framework to deal with know-how about how to find web services to operationalise image analysis pipelines.. We aggregate all the specifications captured into fragments in order to promote among the community members both sharing of image analysis pipeline specifications and cross fertilisation of know-how about how to search for web services. Fragments are represented by rules implemented in the SPARQL language which provides a unified framework to represent both concrete and abstract rules through the CONSTRUCT query form. We then rely on a semantic search engine like CORESE for both backward chaining on the knowledge base of rules and matching with the knowledge base of RDF annotations describing available web services.

Beyond a mix between existing intentional requirement modeling techniques and web semantic models and techniques, our main contribution consists in: (i) providing reasoning and query capabilities for intentional requirement modeling, (ii) leveraging domain knowledge from computer sciences related aspects to neurosciences related aspects in an integrated way and (iii) providing means to support contextualised web services retrieval.

Future works will focus on composition and evolution concerns as well as access rights management, by providing dedicated operators in each step of our approach, in addition to mappings from one step to the other. We also plan to
exploit traceability and reasoning capabilities of CORESE engine to improve responsibility considerations handling during the web services search process. Last, practical work will consist in testing the consistency of our approach through several case studies.